\documentclass[12pt]{article}

\usepackage[english]{babel}
\usepackage{amsmath}
\usepackage{amssymb}
\usepackage{amsbsy}
\usepackage{amstext}
\usepackage{graphicx}

\newcommand{\be}{\begin{eqnarray}}
\newcommand{\ee}{\end{eqnarray}}
\newcommand{\bdm}{\begin{displaymath}}
\newcommand{\edm}{\end{displaymath}}
\newcommand{\ds}{\displaystyle}   
\newcommand{\ba}{\begin{array}}
\newcommand{\ea}{\end{array}}
\newcommand{\pa}[1]{\left(#1\right)}
\newcommand{\paq}[1]{\left[#1\right]}
\newcommand{\az}{\mathcal{S}}
\newcommand{\rr}{\mathcal{R}}
\newcommand{\dpa}{\partial}
\newcommand{\lag}{\mathcal{L}}

\begin{document}

\title{Effective field theory analysis of the self-interacting chameleon}
\author{Hillary Sanctuary and Riccardo Sturani\\
\small Department of Physics, University of Geneva, CH-1211 Geneva,
Switzerland\\
\small e-mail: Hillary.Sanctuary, Riccardo.Sturani@unige.ch}

\maketitle

\begin{abstract}
We analyse the phenomenology of a self-interacting scalar field in the  
context of the chameleon scenario originally proposed by Khoury and Weltman.
In the absence of self-interactions, this type of scalar field can mediate long
range interactions and simultaneously evade constraints from violation of the 
weak equivalence principle. 
By applying to such a scalar field the effective field theory method proposed
for Einstein gravity by Goldberger and Rothstein, we give a thorough 
perturbative evaluation of the importance of non-derivative 
self-interactions in determining 
the strength of the chameleon mediated force in the case of orbital motion.
The self-interactions are potentially dangerous as they can change the
long range behaviour of the field.  Nevertheless, we show that they do not lead to 
any dramatic phenomenological consequence with respect to the linear case and solar system constraints are fulfilled.
\end{abstract}


\section{Introduction}
Despite the well established observation that gravitation and electromagnetism 
are the only long-range forces, the existence of a spin-0 field propagating an 
additional long-range force was conjectured long ago 
\cite{Bergmann:1968ve,Wagoner:1970vr}. Since there is no experimental evidence 
for such a ``fifth'' force, there are basically two ways for a scalar particle 
to exist while maintaining the absence of an extra long-range force.  
This scalar field would have to be either weakly coupled to matter, or massive 
enough for its interactions to be effectively short-ranged, since allowing a
massless scalar field to couple to ordinary matter with gravitational strength
would lead to a stark violation of the \emph{equivalence principle}. The weak 
equivalence principle (WEP) is said to be violated if the coupling depends 
on parameters other than the mass of the particle to which it couples.  
The WEP states that \cite{Will:2008li}
\begin{quote}
\emph{If an uncharged test body is placed at an initial event in
  space-time and given an initial velocity there, then its subsequent
  trajectory will be independent of its internal structure and composition.}
\end{quote}

The standard way to maintain a long-ranged force due to an extra
scalar field is to suppress its coupling to matter.  
The task is not 
straightforward, however, as scalar fields tend to couple to matter with 
gravitational strength, see \cite{Damour:1994zq,Anderson:1997un} for ideas 
based on symmetry principles to suppress scalar couplings.  Otherwise, 
the presence of light scalar fields, or \emph{moduli}, with run-away 
potentials is ubiquitous in super-gravity constructions \cite{Brustein:1992nk}.
Their stabilisation (i.e. by giving them a mass) turns out to be a formidable 
task and has only been achieved in specific constructions \cite{Kachru:2002he}.

A novel mechanism was suggested \cite{Khoury:2003aq,Khoury:2003rn} to suppress 
long-range interactions mediated by a scalar field while allowing its 
fundamental dimensionless coupling to ordinary matter be of order unity. 
This mechanism consists of endowing the scalar field with a mass that 
depends on local matter density. Such a scalar field is called the 
\emph{chameleon} for its ability to evade detection by changing its aspect in 
different environments.  In theory, the chameleon can acquire a large enough 
mass in the neighborhood of the laboratory, allowing it to escape detection and 
searches for violation of the equivalence principle. It remains effectively 
massless in almost empty space, for instance in interplanetary space, so that 
it can propagate to astrophysical or cosmological distances.  
It turns out that the chameleon violates the WEP.

The equivalence principle has been accurately tested. Experiments 
testing differential relative acceleration of massive objects like the earth 
and the moon towards the sun verify the equivalence principle to less than $10^{-13}$ \cite{Will:2008li}.   
Better limits still are expected from forthcoming satellite experiments
\cite{sat}. 

Moreover, precision tests of gravity provide strong constraints on 
the post-Newtonian parameters that are used to
measure possible deviations from standard General Relativity, which
translate into constraints for \emph{scalar-tensor} theories of gravity
and the corresponding \emph{linear} coupling of the extra scalars to matter.
For the Cassini experiment in particular, 
\cite{Bertotti:2003rm} measurement of the Shapiro time-delay in the solar system gives a constraint
on the post-Newtonian parameter $\gamma$ which provides the strongest bound \cite{Will:2008li} 
on scalar-tensor theories endowed with the chameleon mechanism.  For this reason, we consider the chameleon model for bodies with weak self-gravity (unlike neutron stars and black holes).

Several models of modified gravity can actually be rewritten in terms 
of scalar-tensor theories \cite{Maedabook}, for which the same
strict experimental bounds exist.  This is the case for DGP models
\cite{Dvali:2000hr}, massive gravity \cite{massive}, and $f(R)$ theories of 
gravity, where the latter require the chameleon mechanism in order to be 
phenomenologically viable \cite{FofR}.

The chameleon was originally suggested in a cosmological context and has been 
exploited as a quintessential field to give a description of dark energy, see 
\cite{quint} for original suggestions to model dark energy with a scalar field
rolling down a flat potential. In this article, however, we do not wish to 
elaborate on the cosmological implications of the chameleon, but
rather point out a theoretical issue that is phenomenologically relevant for 
the chameleon whenever third or higher order \emph{non-derivative} 
self-interactions of scalar fields are
taken into account.  Self-interactions were already considered, either by 
taking into account the $\phi^4$ self-interaction 
\cite{Feldman:2006wg,Gubser:2004uf}  (where $\phi$ denotes the scalar field)
or by considering a generic type of potential \cite{Mota:2006ed,Mota:2006fz}.  
We show that qualitatively different results are possible if $\phi^3$ 
interactions are taken into account, as derived in \cite{Porto:2007pw}, even if
such modifications are quantitatively negligible for most ranges of parameter 
values. Our analysis is an application of the effective field theory approach
originally proposed in \cite{Goldberger:2004jt} for Einstein gravity.

Gravitational and scalar self-interactions have fundamentally different 
behaviour.  Let us first consider the familiar case of Einstein gravity.  The 
corresponding Newtonian limit is known to work well for most systems of 
interest.  Effects due to graviton self-interactions are suppressed because of 
the specific form of the self-coupling of the graviton. The three-graviton 
vertex, for instance, is of the form $\partial^2 h^3$ (where $h$ denotes the 
graviton with generic polarisation indices and $\dpa$ a generic spacetime 
derivative) and the corresponding correction to
the Newtonian potential is proportional to $1/r^2$ \cite{Goldberger:2004jt}.  
The correction is thus suppressed with respect to the leading Newtonian 
contribution by a factor $r_s/r$ (where $r_s$ is the Schwarzchild radius of the
massive object giving rise to the gravitational field).  These corrections are 
indeed negligible for objects that are sufficiently distant or diffuse.  

The scenario is different for the case of scalar self-interactions, as a
scalar field ca have a \emph{non-derivative} $\phi^3$ self-coupling.  
Two powers less of momentum result in
two powers more of $r$ in the self-interaction amplitude, or the effective 
potential, which in turn leads to a \emph{logarithmic} correction to the 
``Newtonian'' $1/r$ potential.  This next-to-leading order correction 
overcomes the Newtonian potential at large enough distances.  So $\phi^3$ 
self-interactions may significantly change the long-range behaviour of the 
scalar field, in contrast to the gravitational case.  This argument is explained
quantitatively in sec.~\ref{self_int}.

Field theory techniques like Feynman diagrams were first applied to gravity 
and scalar tensor theories in \cite{Damour:1995kt}, where \emph{derivative} 
self-interactions of the type $\partial^2\phi^n$ (for $n=3,4$) were considered. 
Derivative self interactions were also studied in other models 
equivalent to tensor scalar theories, such as the DGP and 
massive gravity models mentionned above, where the leading-order extra-scalar
self-interactions are of the type $\partial^n\phi^3$, with $n=4,6$
respectively.

This article is organised as follows.  In sec.~\ref{cham_overview}, we recall the
basic ingredients and known result of chameleon models. In sec.~\ref{self_int},
we study the effect of self-interactions on the chameleon mediated potential 
for a generic choice of parameters. We show that the self-interactions may \emph{grow} over 
astrophysical distances, even though for realistic values of the physical 
parameters, self-interactions do not lead to any dramatic phenomenological 
consequences. We finally conclude in sec.~\ref{conclusion}.

\section{Chameleon in brief}
\label{cham_overview}
Consider the following action
\be
\ba{rl}
\az=&\ds\int d^4 x\sqrt{-g}
\paq{\frac \rr{16\pi G_N}-\frac 12\dpa_\mu\phi\dpa^\mu\phi-V(\phi)}+\\
&\ds\int d^4x\sqrt{-g} \lag_{matter}\pa{\psi^{(i)},g^{(i)}_{\mu\nu}}\,,
\ea
\ee
with signature $(-,+,+,+)$ for the metric $g_{\mu\nu}$, 
$g={\rm det}g_{\mu\nu}$, $\rr$ the Ricci scalar, $G_N$ the Newton constant, 
$\phi$ the chameleon field subject to the fundamental potential $V(\phi)$ and 
coupled to matter fields $\psi^{(i)}$ via the modified metric 
$g^{(i)}_{\mu\nu}\equiv e^{\beta_i\phi/{M_{Pl}}}g_{\mu\nu}$.  We also define 
$M_{Pl}\equiv \pa{8\pi G_N}^{-1/2}\simeq 2.4\cdot 10^{18}$GeV.  
 
Let the fundamental potential be an inverse power-law of the form
\be
V(\phi)=M^4\pa{M/\phi}^{\alpha}\,,
\ee
with $\alpha$ positive and $M$ the fundamental mass scale of the problem 
(possibly much smaller than $M_{Pl}$).  The specific form of the potential, 
however, is not crucial for the result we wish to discuss, but we nevertheless 
require it to be a decreasing function without a minimum.  The chameleon 
couples to the trace of the energy momentum tensor
of matter $T^{(i)}\equiv g^{\mu\nu}T^{(i)}_{\mu\nu}$.   For non relativistic 
matter, we can safely replace $T^{(i)}\simeq -\rho$, where $\rho$ is the energy
density in the $i$-th particle species.  We consider only one particle species 
and henceforth drop the index $i$. This coupling induces an effective potential
\be
V_{eff}(\phi)=V(\phi)+\rho e^{\beta\phi/M_{Pl}}=
M^4\pa{\frac M\phi}^\alpha+\rho e^{\beta\phi/M_{Pl}}\,.
\ee
Use of such an inverse power-law as the fundamental potential is an assumption,
but qualitatively similar potentials appear in supergravity compactifications 
inspired by superstring constructions \cite{Grana:2005jc}.
The minimum of the effective potential, defined by $V_{eff,\phi}(\bar\phi)=0$, 
is approximately given by 
\be
\bar\phi\simeq M \pa{\frac{\alpha}{\beta}\frac{M_{Pl}M^3}{\rho}}^{1/(\alpha+1)}
\,.
\ee
We consider a range of parameters for which this formula applies.  
In particular, for this formula to hold $\bar\phi$ must be sub-Planckian,
which is the case if $M$ is small enough and $\rho$ is large enough, i.e.
\be
\label{mrho}
M<\pa{\frac{\beta^\alpha}\alpha M_{Pl}^\alpha\,\rho}^{\frac 1{\alpha+4}}\,.
\ee
For instance, $\rho=1\rm{g/cm^3}\simeq\pa{5\cdot 10^{-2}MeV}^4$, $\alpha=2$ 
relation (\ref{mrho}) implies $M<3$ TeV.

The effective potential, Taylor expanded around the minimum, is
\be 
\label{veff}
\ba{rl}
\ds V_{eff}(\phi) = \frac{m_\phi^2}{2}\pa{\phi-\bar\phi}^2+
\frac{g_3M}{3!}\pa{\phi-\bar\phi}^3+&
\ds\frac{\lambda}{4!}\pa{\phi-\bar\phi}^4+\\
&\ds\sum_{n>4}\frac{g_n}{n!M^{n-4}}\pa{\phi-\bar\phi}^n\,,
\ea
\ee
where the following $\rho$-dependent parameters have been defined: the mass 
$m_\phi$, the dimensionless trilinear coupling $g_3$, the quartic coupling 
$g_4\equiv\lambda$ and the generic dimensionless n-linear coupling $g_n$.  
Such parameters are related to derivatives of the effective potential 
evaluated at $\bar\phi$ and they are approximately given by
\be
\ba{rcl}
m_\phi^2 &\simeq& \ds
\pa{\alpha+1}M^2\pa{\frac{\beta \rho}{M^3 M_{Pl}}}^{\frac{\alpha+2}{\alpha+1}} 
\exp(\beta \bar\phi/M_{Pl})\,, \\
g_3 &\simeq& \ds -\pa{\alpha+1}\pa{\alpha+2} \pa{
\frac{\beta \rho}{M^3M_{Pl}}}^{\frac{\alpha+3}{\alpha+1}} 
\exp(\beta \bar\phi/M_{Pl})\,, \\
\lambda &\simeq& \ds (\alpha+1)(\alpha+2)(\alpha+3)\pa{
\frac{\beta \rho}{M^3 M_{Pl}}}^{\frac{\alpha+4}{\alpha+1}}
\exp(\beta\bar\phi/M_{Pl})\,.
\ea
\ee

These parameters indeed depend on the local matter density.  
Setting $M=1$ eV and $\alpha=\beta=1$, typical values for the earth, 
the atmosphere and interplanetary space are summarised in tab.~\ref{data}. 

\begin{table}
\begin{tabular}{|c|c|c|c|c|c|c|}
\hline
& $\rho {\rm (g/cm^3)}$ & $m_\phi {(\rm eV)}$ & $\lambda_C$ (m) & $-g_3$ & 
$\lambda$ & $\bar\phi/M$\\
\hline
Earth & $5.5$ & $1.4\times 10^{-6}$ & $0.14$ & $6.4\times 10^{-16}$ & 
$2.6 \times 10^{-19}$ & $9.8\times 10^3$\\
\hline
Atmosphere & $1.3 \times 10^{-3}$ & $2.8 \times 10^{-9}$ & 
$ 72 $ & $3.6 \times 10^{-23} $  & $  2.2 \times 10^{-28} $ &
$6.4\times 10^5$\\
\hline
Interplanetary Space & $10^{-24}$ & $4 \times 10^{-25}$ & $5\times 10^{17}$ &
$2 \times 10^{-65}$ & $ 4 \times 10^{-81}$ & $2\times 10^{16}$\\
\hline
\end{tabular}
\caption{Value of the parameters for different densities ($\alpha=\beta=1$, 
$M=1$eV). For comparison, $1 \rm{pc}\simeq 3\times 10^{16}$m.}
\label{data}
\end{table}

For later reference $g_{n}$ is given by
\be
g_{n}=(-1)^{n} \frac{(\alpha+n-1)!}{\alpha!}
\pa{\frac{M}{\bar\phi}}^{\alpha+n}\,,
\ee
where the factor involving the exponential of $\phi$ has been neglected.
We will see in sec.~\ref{self_int} that the interactions with $n \geq 5$ are of
negligible phenomenological impact.

Following \cite{Khoury:2003rn,Mota:2006fz}, we determine the 
profile of the chameleon in the case of a spherically symmetric source 
of density $\rho_c$ and radius $R$, surrounded by an environment of density 
$\rho_\infty$.  As a first approximation of the long range behaviour of the 
chameleon in the presence of massive objects, one can linearise the 
time-independent, spherically symmetric equation of motion by keeping only the 
quadratic term in the potential expansion (\ref{veff}), to obtain
\be
\label{lin_eq}
\phi''+\frac 2r\phi'-m^2_c(\phi-\phi_c)&=&0\,,
\ee
where $m_c,\phi_c$ respectively denote the value of $m_\phi,\bar\phi$ inside 
the source and we have neglected the curvature of the space-time induced 
by the source.
This approximation is usually referred as the
\emph{``thin shell''} approximation in the literature. In the $m_c R\gg 1$ 
case, the ``thin shell'' approximation 
is appropriate, since $\phi$ is close to the minimum within most of the 
source and only becomes significant for a thin shell close to its edge,
as discussed below.\\
Following standard notation, we introduce $\phi_\infty$, the field value at 
the minimum of the effective potential for $\rho=\rho_\infty$ (outside the 
source).  In particular, 
we consider the case in which the environment, outside the source, is 
endowed with such a small density that the Compton wavelength of the 
corresponding chameleon is much greater than the length scales of interest ($m_\infty r\ll 1$).

The solution to the linearised equation (\ref{lin_eq}) inside the source
object, and the corresponding solution outside it, is 
\renewcommand{\arraystretch}{1.5}
\be
\label{thin_shell}
\ba{rcl}
\ds \phi(r<R)&=& \ds A\frac{\sinh(m_c r)}r+\phi_c\,,\\
\ds \phi(r>R)&=&\ds B\frac{e^{-m_\infty (r-R)}}r+\phi_\infty\,,
\ea
\ee
\renewcommand{\arraystretch}{1}
where the integration constants are fixed by requiring the solution and its 
derivative to continuously match at the boundary $r=R$, giving
\renewcommand{\arraystretch}{2}
\be
\label{coeffs}
\ba{rcl}
A &=& \ds \paq{\frac{1+m_\infty R}{m_c R\coth(m_c R)+m_\infty R}}
\frac{R\pa{\phi_\infty-\phi_c}}{\sinh(m_c R)}\,,\\
B &=& \ds \paq{\frac{m_cR\coth(m_cR)-1}{m_cR\coth(m_cR)+m_\infty R}}
R(\phi_c-\phi_\infty)\,.
\ea
\ee
\renewcommand{\arraystretch}{1}
Other solutions can be obtained as in \cite{Khoury:2003rn,Mota:2006fz} using 
different approximations, but here eq.~(\ref{thin_shell}) encompasses the 
relevant physics for macroscopic orbiting bodies, like 
planets and stars.\\
For instance in the case $\phi(r)>\phi_c$ inside the source 
(i.e. $r<R$), the effective potential can be approximated by its increasing 
branch leading to the approximate equation of motion 
\be
\ds\phi''+\frac 2r\phi'+\frac{\beta\rho}{M_{Pl}}=0\,,
\ee
and retaining the approximate form (\ref{lin_eq}) of the equation of motion 
\emph{outside} the source of total mass $M_c$, one has the solution
\be \label{whole}
\ba{rcl}
\ds \phi(r<R)&=&\ds -\frac{\beta\rho_c r^2}{6 M_{Pl}}+{\rm const}\,, \\
\phi(r>R) &=& \ds \frac{\beta}{4\pi M_{Pl}}\frac{M_ce^{-m_\infty(r-R)}}r+
\phi_\infty\,.
\ea
\ee
This profile is a self-consistent solution for $\phi(r<R)>\phi_c$ and has been 
named the \emph{thick shell} solution, occurring for $m_cR<1$.

For objects of astrophysical 
interest, like planets and stars, the size of the source of 
the chameleon field is larger than its corresponding Compton wavelength,
as can be checked from tab.~\ref{data}.
Consequently we focus on eqs.~(\ref{thin_shell}), in which case only a 
\emph{thin shell} at the surface of the object contributes to the overall 
chameleon field. In this case the solution outside the source can be 
rewritten as 
\be
\label{phi_pot}
\phi(r>R)\simeq -\frac{\beta_{eff}}{4\pi M_{Pl}}\frac{M_ce^{-m_\infty(r-R)}}r
+\phi_\infty\qquad m_cR\gg 1\,,
\ee
where
\be
\label{b_eff}
\beta_{eff}=
\frac{3\phi_\infty M_{Pl}}{\rho_cR^2}\,.
\ee  

The standard Newtonian potential is analogous to eq.(\ref{phi_pot}), with 
$\beta_{eff}$ replaced by unity, thus the suppression of the chameleon
coupling is conveniently parameterized by $\beta_{eff}$. 
For instance, considering the earth as a sphere of radius $6.4\times 10^6$m
with homogeneous density $\rho_\oplus\simeq 5.5 \rm g/cm^3$ immersed in the 
galactic medium made of dark matter and baryonic gas with density 
$\rho_G\simeq 10^{-24} \rm g/cm^3$, then
$\beta_{eff}\simeq 5\times 10^{-3}$ for $M=1$eV, and $\alpha=\beta=1$.

The thin-shell solution has the beneficial effect of suppressing the otherwise 
phenomenologically dangerous coupling to matter, but it does so in a 
non-universal manner.  The chameleon couples to matter not only through the 
total mass of the object, but also through parameters such as the radius of the
source.  This peculiar coupling leads to violation of the equivalence 
principle, even in its weak form.

There is nevertheless one major caveat in the above analysis in obtaining the 
solutions, due to the approximations that have been made. 
In the thin-shell case, 
the effective potential is approximated by its quadratic 
expansion around the minimum $V_{eff}(\phi)\simeq m^2(\phi-\phi_c)^2/2$.  
The effect of higher order terms on the solution outside the source 
should be checked, however, since they may be important for the lowest order 
solution (\ref{thin_shell}) at the boundary of the source. 
 Naively substituting the solution (\ref{thin_shell}) into the potential 
expansion (\ref{veff}), one realizes that the linear term in the equation of 
motion (the mass term) is dominated by the tri-linear and quartic interaction 
term for 
$r\lesssim \frac{1}{2}(\alpha+2)R$ and 
$r \lesssim \sqrt{\frac{1}{3!}(\alpha+2)(\alpha+3)}R$, respectively, which are 
both clearly outside the source. In general, the self-interaction terms
$\phi^n$ in the expansion (\ref{veff}) for $n\ge 2$ dominate at distances 
\be 
\label{rstar}
r\lesssim 
\pa{\frac{1}{\pa{n-1}!} \frac{(\alpha+n-1)!}{(\alpha+1)!}}^{\frac 1{n-2}}R\,.
\ee 
The fact that all of the higher order terms overcome the linear term indicates 
the need for a more thorough analysis.

In \cite{Feldman:2006wg}, the effect of a $\lambda\phi^4$ interaction is 
studied using approximate analytic methods. The resulting profile for the 
chameleon in the thin shell case is still $1/r$ outside the source $r>R$ (and
within the Compton wavelength of the chameleon $m_\infty r \ll 1$), but the 
effective coupling reduces to

\be
\beta_{eff}\simeq \frac{M_{Pl}}{M} \lambda^{-1/2}\qquad 
{\rm if} \qquad \lambda> \pa{\frac{M_{Pl}}{\beta M}}^2\,, 
\ee
which occurs for large enough $\lambda$.
This is a non-perturbative effect which cannot be reproduced in a 
perturbative analysis. 
In \cite{Mota:2006fz} an approximate analytic analysis of the non-linear
regime is performed, and a result equivalent to (\ref{phi_pot}) is obtained
(see next to last eq. in sec. IIID of \cite{Mota:2006fz}). They show that
the full form of the potential does not substantially alter the solution 
close to the surface of the source, where the matching between the
inner and the outer solutions in (\ref{thin_shell}) is made.
The argument of \cite{Mota:2006fz} is roughly as follows. Since, at large
distances, the solution can only be of the type of the second equation in 
(\ref{thin_shell}) and since $\phi(r)>0$, we must have $-\phi_\infty R<B<0$ 
so that
$\phi(r)>\phi_{critical}(r)\equiv\phi_\infty\pa{1-Re^{-m_\infty(r-R)}/r}$. 
The limiting value $\phi_{critical}(r)$ coincides with the approximate 
solution (\ref{phi_pot}). If, on the other hand, 
$\phi(r)\gg \phi_{critical}(r)$ outside of the body, then one would 
simply be in the case leading to solution (\ref{whole}).

In sec.~\ref{self_int} we use a perturbative effective field theory method 
borrowed from \cite{Goldberger:2004jt} which gives systematic estimates of 
scalar field self-interactions, particularly relevant to determine the 
corrections to the chameleon profile outside a source where the 
potential is very shallow. We focus on the \emph{thin shell} solutions
as it is the astrophysically relevant solution.

\section{Corrections to the potential from chameleon self-interactions}
\label{self_int}
We have so far recalled how the \emph{thin-shell} solution is 
obtained and mentioned its corrections treated in the literature.  
We present here another tool to compute the chameleon profile beyond the 
linearised approximation. We adopt an effective approach in which sources
are considered to be point-like and coupled to a massive chameleon with
strength $\beta_{eff}$, giving the original chameleon-mediated potential 
(\ref{phi_pot}).
In addition to the mass term, the chameleon field is subject to the 
effective potential $V_{eff}$ of eq.(\ref{veff}). To simplify the problem
we initially truncate the effective potential $V_{eff}$ to cubic order. The 
effect of higher order interactions in $V_{eff}$ are considered later in 
this section. 

We study the following Lagrangian
\be
\label{ac_phi}
S=-\frac 12\int d^4x \sqrt{-g}\,\pa{\dpa_\mu\phi\dpa^\mu\phi+m^2_\phi\phi^2+
\frac{g_3M}3\phi^3}+\beta_{eff}\frac{M_c}{M_{Pl}}\phi \int dt\,.
\ee 
The $\phi$ field is redefined so that its minimum is at $\phi=0$ and 
a constant term in the Lagrangian is neglected.
With $g_3=0$, this Lagrangian reproduces precisely the potential of the thin 
shell solution (\ref{phi_pot}) outside the source. 
Following \cite{Goldberger:2004jt}, where effective field theory methods for 
gravity are discussed, the problem is treated non-relativistically, thus
splitting the kinetic term of (\ref{ac_phi}) into
\be
\dpa_\mu\phi\dpa^\mu\phi=\delta_{ij}\dpa_i\phi\dpa_j\phi-\dot\phi^2\,,
\ee
and treating the time derivative as an interaction term.
In terms of Fourier transformed functions
\be
\phi_{\bf k}(t)\equiv\int \frac{d^3x}{\pa{2\pi}^3}\phi(t,{\bf x})
e^{i{\bf k}\cdot{\bf \cdot x}}\,,
\ee
for the $\phi$-propagator we have
\be
\langle\phi_{\bf q}(t)\phi_{\bf k}(0)\rangle=\pa{2\pi}^3
\delta^{(3)}({\bf q}+{\bf k})\delta(t)\frac 1{k^2+m_\phi^2}\,,
\ee
where the four-momentum $k^\mu=(k^0,{\bf k})$, with 
$k\equiv\sqrt{{\bf k}\cdot {\bf k}}$.

\begin{figure}
\centering
\includegraphics[width=.6\linewidth]{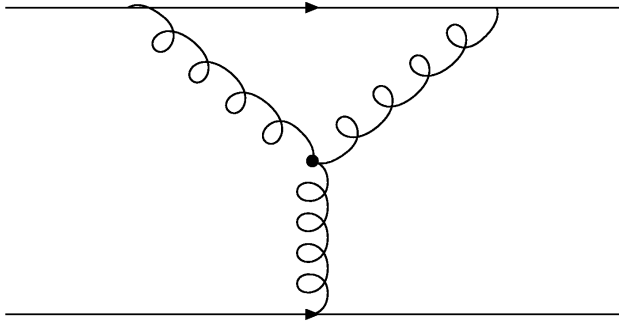}
\caption{Feynman diagram displaying the contribution to the effective 
potential proportional to the tri-linear interaction $g_3M\phi^3$.}
\label{3_int}
\end{figure}

To account for the effect of the tri-linear interaction on the chameleon 
potential, the amplitude represented by the Feynman diagram in 
fig.~\ref{3_int} has to be computed. 

The source masses are $M_{c1},M_{c2}$ with trajectories ${\bf x_1}(t),{\bf x_2}(t)$ respectively, 
see \cite{Porto:2007pw}. The 3-point function of 
the $\phi$-field is
\be
\ba{rl}
\ds\langle T(\phi(x_1)\phi(x_2)\phi(x_3))\rangle&\ds =3!(-ig_3M)
\delta(t_1-t_2)\delta(t_2-t_3)\times\\
&\!\!\!\!\!\!\!\!\!\!\!\!\!\ds \int\prod_{r=1}^3\frac{d^3k_r}{\pa{2\pi}^3}
e^{i{\bf k}_r\cdot {\bf x}_r}\ \pa{2\pi}^3\delta^{(3)}
\pa{\sum_{r=1}^3 {\bf k}_r}\prod_{r=1}^3\frac{-i}{k_r^2+m^2_\phi}\,.
\ea
\ee
Then the diagram in fig.~(\ref{3_int}) can then be computed to give a 
contribution to the effective action \cite{Porto:2007pw}
\renewcommand{\arraystretch}{1.6}
\be
\label{3phi}
\ba{ccl}
\mbox{fig.~\ref{3_int}} &=& \ds \frac 12\pa{\frac{-iM_{c1}}{M_{Pl}}}^2
\pa{\frac{-iM_{c2}}{M_{Pl}}}\int dt_1dt_1'dt_2
\langle T\pa{\phi(x_1)\phi(x_1')\phi(x_2)}\rangle\\
&=&\ds - i g_3 \frac{\beta_{eff}^3MM_{c1}^2M_{c2}}{M_{Pl}^3}
~\int dt\log (m_\phi |{\bf x}_1-{\bf x}_2|)\,.
\ea
\ee
\renewcommand{\arraystretch}{1}
In the computation, the mass term in the propagator is neglected, as we 
consider distances (here and in the rest of this section) $r\ll 1/m_\phi$, 
but it is reinserted at the end of (\ref{3phi}) as an infrared regulator. 
One immediately notices that such a 
contribution to the effective potential \emph{grows} with distance, which it eventually overcomes the lowest order solution 
(\ref{thin_shell}) 
is a distance at $r_*$.

When the body is taken to be the earth, this distance corresponds to  
\be
r_*\simeq \frac 1{g_3} \frac{M_{Pl}}{\beta_{eff}M_cM}\,,
\ee
i.e. $r_*\simeq 10^6$ Mpc (larger the the Hubble radius) for $M=1$eV
and $R=R_\oplus$, $\rho=\rho_\oplus$, so negligible for reasonable 
distances.

\begin{figure}[t]
\centering 
\includegraphics[width=0.6\linewidth]{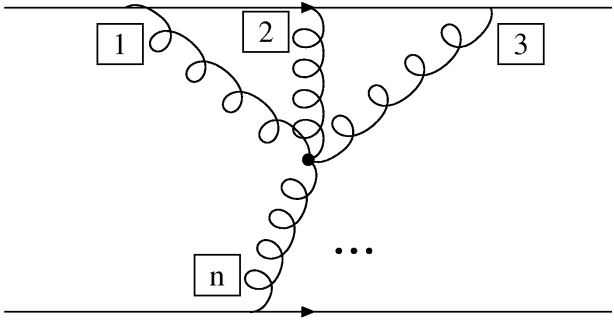}
\caption{Feynman diagram displaying the generic contribution to the 
effective potential proportional to the n-point interaction 
$g_nM^{4-n}\phi^n$.}
\label{n_int}
\end{figure}

For higher order interactions, it is not necessary to perform the actual 
computation of the relevant Feynman diagram fig. \ref{n_int}.
Using effective field theory methods, we can indeed estimate the scaling of the relative amplitude for the contribution to the effective potential 
mediated by all the other $g_nM^4\pa{\phi/M}^n$ interactions.

In effective field theory, one usually has an expansion in a perturbative 
parameter. For instance, in general relativity, post-Newtonian computations 
for two bodies gravitating around each other have the typical 
velocity $v$ of the system as expansion parameter (n-th PN computation means 
$v^{2n}$ correction to the Newtonian result). In our case, as we will see, we 
have a proliferation of dimensionless scales, so it is not possible to 
identify a single expansion parameter, however the method of 
\cite{Goldberger:2004jt} can still be applied to assess the strength of 
different contributions to the effective potential.

The effective action is computed perturbatively by applying systematic power 
counting rules, relying on the basic assumption that Newtonian gravity is
still responsible for the leading interaction, so that the virial relation 
$v^2\sim G_NM_c/r$ holds. Each insertion on the world-line of a source, 
propagator, $\phi$ vertex, requires one of the following factors 
\begin{itemize}
\item $\beta_{eff}dt d^3k M_c/M_{Pl}$ for a particle-chameleon vertex
\item $g_nM^{4-n}dt\,(d^3k)^n\,\delta^{(3)}(k)$ for each n-$\phi$ vertex
\item $\delta(t)\delta^{(3)}(k)\times 1/k^2$ for each propagator. 
\end{itemize}
In standard Einstein gravity, two gravitating bodies exchange gravitons 
mediating the gravitational potential with momenta $k^\mu=(k_0,{\bf k})$, 
where $k^0\sim v/r$ and $k\sim 1/r$ \cite{Goldberger:2004jt}. 
One can then assign the scaling $x^0\sim r/v$, 
$x\equiv |{\bf x}|\sim r$, $k\sim 1/r$ and consequently
$\delta^{(3)}(k)\sim r^3$, $d^3k\sim 1/r^3$, $\delta(t)\sim v/r$, $dt\sim r/v$.

These rules can be applied to a diagram contributing to 
the effective potential between two massive objects due to the exchange of 
$\phi$-fields and involving an n-$\phi$ vertex ($n>2$), so the scaling becomes
\be
\label{scaling_ch}
fig.~\ref{n_int}\sim 
g_n\beta^n_{eff} M^{4-n}\pa{\frac{M_c}{M_{Pl}}}^n\frac{r^{4-n}}v\,.
\ee
We recall that the simple one graviton exchange scales as 
\cite{Goldberger:2004jt}
\be
\label{ldef}
\pa{\frac{M_c}{M_{Pl}}}^2 \frac tr\simeq M_cvr\equiv L\,,
\ee
where the virial relation $M_c/(M_{Pl}^2r)\sim v^2$ has been used.  
Using both (\ref{ldef}) and the same virial relation in the chameleon 
amplitude scaling (\ref{scaling_ch}) one obtains
\be
\label{scaling}
fig.~\ref{n_int}\sim 
g_n\beta_{eff}^n L \pa{\frac{r_s}{l_{Pl}}}^2 \pa{\frac{r_s}{r}}^{n-4}
\pa{\frac{M}{M_{Pl}}}^{4-n}\,,
\ee
where $r_s\equiv 2G_NM_c$ is the Schwarzchild radius of the source and 
the Planck length $l_{Pl}\equiv M_{Pl}^{-1}$ has been introduced.  
The proliferation of dimensionless ratios ($r_s/r$, $M/M_{Pl}$ and 
$M_c/M_{Pl}$) renders this result less immediate to interpret, but the 
substitution of actual values for physical parameters provides insight as to 
what the scaling (\ref{scaling}) means. Let us observe that $L$ is 
the scaling of the action involving the Newtonian potential 
($\sim dt\,G_NM_c^2/r$), and $\beta_{eff}^2 L$ is the scaling of the 
contribution to the effective action obtained by a diagram analogous to 
fig.~\ref{n_int} but with just one $\phi$-propagator. 
Apart from $\beta_{eff}^2L$, there are in eq.(\ref{scaling}) extra terms 
involving the ratios $r_s/r<1$, $M/M_{Pl}<1$ and $r_s/l_{Pl}>1$. 
To evaluate the importance of the diagram involving the 
n-$\phi$ vertex, like in fig.~\ref{n_int} for the case of two orbiting bodies, 
an estimation of the parameters is necessary.
Numerical values are summarised by
\renewcommand{\arraystretch}{1.5}
\be
\label{estimate}
\ba{ccl}
\ds\pa{\frac{r_s}{l_{Pl}}}^2 &\simeq & \ds \pa{4\cdot 10^{37}\times 
\frac{M_c}{M_\odot}}^2\,,\\
\ds \frac{r_s}{r} &\simeq & \ds 2\cdot 10^{-8} \times
\frac{M_c}{M_{\odot}}\pa{\frac r{1 {\rm AU}}}^{-1}\,,\\
\ds \frac{M}{M_{Pl}} &\simeq &\ds 4\cdot 10^{-28}\times \frac{M}{1eV}\,,\\
\ds \beta_{eff}& \simeq & \ds 2.2 \cdot 10^{-6} \times 
\pa{\frac{\phi_{\infty}}{10^{16}{\rm eV}}}\pa{\frac{R}{R_\odot}}
\pa{\frac {M_c}{M_{\odot}}}^{-1}\,.
\ea
\ee
\renewcommand{\arraystretch}{1}
For reasonable values of the parameters, such corrections are less and less 
relevant as $n$ increases, and the resulting contribution to the two-body 
potential goes with distance as $1/r^{n-3}$.\\
The case of the amplitude in eq.~(\ref{3phi}) corresponds to $n=3$, i.e.
a logarithmic potential, or as eq.(\ref{scaling}) shows, a potential whose 
$r$ dependence displays one power more than the $1/r$, Newtonian, usual 
behaviour of single particle exchange. 
For $n=4$ the amplitude is $a$ times the contribution from 
the diagram with a single chameleon exchange, where $a$ is given by
\be
a\equiv\beta_{eff}^2\lambda \pa{r_s/l_{Pl}}^2\,.
\ee
Taking the value for $\lambda$ in interplanetary medium from tab.~\ref{data},
we see that $a$ becomes a strong suppression factor. 
To answer the issue mentionned above eq.(\ref{rstar}), it is of little 
importance that the higher order terms in the expansion are as 
important as the mass term near the surface of the body: as we are 
considering here the case $r\ll 1/m_\infty$ the potential is simply 
negligible for the determination of the chameleon profile
in the region well within the Compton wavelength, where the Yukawa suppression 
has not yet taken place.\\
Going from a diagram with a $n$-point interaction vertex to one with a $n+1$
boils down to multiplying the amplitude by a factor given by
\be
\label{rel_pot}
\frac{g_{n+1}}{g_n}\beta_{eff}\pa{\frac{r_s}r}\pa{\frac{M_{Pl}}M}\simeq 
\beta_{eff}\pa{\frac{r_s}r}\pa{\frac{M_{Pl}}{\phi_\infty}}\,,
\ee
which again is smaller than unity for reasonable parameters, as can be 
estimated from tab.~\ref{data} and (\ref{estimate}).
The perturbative series is then under full control.

The kinetic term and the potential of the chameleon field, including all 
non-linearities,  contribute to the energy-momentum tensor which, in turn, 
affect the background gravitational field via the Einstein equations.  
It is therefore necessary to verify that this chameleon 
induced backreaction on the gravitational field is negligible compared to the 
background gravitational field.
This verification is done by 
comparing the respective chameleon and gravitational potentials as follows.  
Let $T$ be the generic entry of the energy-momentum tensor of the chameleon 
field and $h$ the metric change due to its effect.   
We then obtain from Einstein equations:
\be
\label{back_reaction}
\ba{rl}
\ds h''\sim 8\pi G_N T\sim \frac{{\phi'}^2+V(\phi)}{M_{Pl}^2}\sim &\ds 
\frac{{\phi'}^2+m_\infty^2\pa{\phi-\phi_\infty}^2}{M_{Pl}^2}\sim\\
&\ds
\beta_{eff}^2\frac{r_s^2}{r^4}+\beta_{eff}^2m_\infty^2\pa{\frac{r_s}{r}}^2\,,
\ea
\ee
where only the contribution of the quadratic part of the chameleon potential 
has been considered. Analogous reasoning behind eq.~(\ref{rel_pot}) leads to 
the conclusion that the contribution from higher order chameleon 
self-interactions are sub-dominant with respect to the 
$m_\infty^2(\phi-\phi_\infty)^2$ term.
For $r<m_\infty^{-1}$, the term proportional to $1/r^2$ is sub-dominant with 
respect to the one proportional to $1/r^4$, thus leading to the estimate 
$h\sim \beta_{eff}^2 r^2_s/r^2$, whose effect is suppressed by a small
factor $\beta^2_{eff}$ compared to the first order Post-Newtonian correction 
to the gravitational potential and a factor $\beta_{eff}r_s/r$ compared
to the leading chameleon solution. 
In the opposite regime $r>m_\infty^{-1}$, the Yukawa suppression takes place.
We have thus verified that the metric backreaction due to the chameleon is 
indeed negligible compared to the background solution.

Extra long-range forces have to fulfill experimental constraints, as Einstein 
gravity is known to work perfectly well. For the chameleon, such constraints
have already been taken into consideration in several articles, see e.g.
\cite{Mota:2006fz,Brax:2007vm}, in the following we quote the ones 
relevant for orbiting bodies in the effective theory defined by 
eq.~(\ref{ac_phi}).\\
A different acceleration of the moon and the earth towards the sun 
$\eta_{\oplus -m}$ would have been detected, had it exceeded the fractional 
value of $\eta_{\oplus -m}^{(exp)}\sim 10^{-13}$
\cite{Will:2008li}.
Violating the WEP, the chameleon mediated force induces a differential 
acceleration $\eta_{\oplus -m}^{(cham)}$ for the earth and the moon toward the 
sun (normalised to the ordinary acceleration towards the sun at $1AU$),
of strength roughly given by
\renewcommand{\arraystretch}{1.5}
\be
\ba{rl}
\ds\eta_{\oplus -m}^{(cham)}\simeq &\ds{\beta_{eff}}_{\odot}
\left|{\beta_{eff}}_\oplus-{\beta_{eff}}_m\right| \simeq 
\\
&\ds 16\pi^2
\frac{\phi_\infty^2 R_\odot R_m}{M_\odot M_m/M_{Pl}^2}
\simeq 3.6\cdot 10^{-7}\times 
\pa{\frac M{1{\rm eV}}}^{2\frac{\alpha +4}{\alpha +1}}\,,
\ea
\ee
\renewcommand{\arraystretch}{1}
which evades the bound on $\eta_{\oplus -m}$ for 
$M\lesssim 3\times 10^{-7\frac{\alpha+1}{2(\alpha+4)}}$ eV (where $\alpha$ and
$\beta$ have been set to 1 elsewhere than in the exponent).

Considering other astrophysical constraints from orbiting bodies,
usually such bounds are expressed in terms of the Brans-Dicke parameter
$\omega_{BD}$, defined by the action $S_{BD}$ ruling the dynamics of the 
Brans-Dicke scalar field $\Phi$,
\be
\label{brans-dicke}
\ba{rl}
\ds \az_{BD}=\az_\Phi+\az_m=\frac{M_{Pl}^2}2&\ds
\int d^4x\sqrt{-g}\paq{e^\Phi\pa{R-\omega_{BD}\pa{\dpa\Phi}^2}}+\\
&\ds\int d^4x\sqrt{-g}\lag_m(\psi^i,g_{\mu\nu})\,.
\ea
\ee
Once $\Phi$ is canonically normalised, i.e. 
$\Phi^c\equiv\Phi/(M_{Pl}\omega_{BD}^{1/2})$, and the metric rescaled by a 
factor $e^{-\Phi}$, eq.~(\ref{brans-dicke}) gives a coupling to matter
\be
\az_m=\frac{M_c}{2M_{Pl}\omega_{BD}^{1/2}}\Phi^c\int dt\,.
\ee
Present bounds from Cassini spacecraft \cite{Bertotti:2003rm}, for instance, 
give 
$\omega_{BD}>4\cdot 10^4$ \cite{Will:2008li}, translating into 
$\beta_{eff}<3\cdot 10^{-3}$ for the chameleon, which is easily evaded
by allowing a small enough $M$ (e.g. $M<$ few eV in the case of the earth).

\section{Conclusions}
\label{conclusion}
We have given a systematic analysis of the corrections to the chameleon
mediated potential due to non-derivative $n-$th order self-interactions, 
performed through the implementation of the effective field theory method 
originally proposed in \cite{Goldberger:2004jt} for General Relativity, in the
case of weakly self interacting bodies (which include stars and planets but 
not black holes).
The analysis of the effects of self interaction is crucial for understanding
if the result for the free field can be extended to the fully interacting case.
Trilinear interactions have been shown to be potentially dangerous, as their
contribution to the potential grows with distance with respect to the 
lowest order effect, but for ordinary values of the parameters at play, such
contributions are actually harmless.
Contributions from higher order interactions decrease faster with distance and
they are also negligible.\\
We have presented a thorough perturbative analysis of such effects with the 
help of effective field theory methods and the powerful tool of Feynman 
diagrams. 
Once the relevant scales in the problem are identified, even if they are 
multiple, as in this case, it is possible to set a perturbative expansion 
which allows for the assessment of different diagrams.

\section*{Acknowledgments}
This work is supported by the Fonds National Suisse. The work of R.S. is 
supported by the Boninchi foundation.
The authors wish to thank M. Maggiore for discussions and support.

\end{document}